# HYDROGEN STORAGE IN CARBON DERIVED FROM SOLID ENDOSPERM OF COCONUT

**Viney Dixit, Ashish Bhatnagar, R. R. Shahi, T. P. Yadav, O. N. Srivastava\***
Hydrogen Energy Center, Department of Physics, Banaras Hindu University,
Varanasi-221005, India
\* Corresponding author: E-mail: heponsphy@gmail.com; Tel: 00915422368468

## Abstract

Carbons are being widely investigated as hydrogen storage material owing to their light weight, fast hydrogen adsorption kinetics and cost effectiveness. However, these materials suffer from low hydrogen storage capacity, particularly at room temperature. The aim of the present study is to develop carbon-based material from natural bio-precursor which shows at least moderate hydrogen storage at room temperature. For this purpose, hydrogenation characteristics of carbon derived from solid endosperm of coconut is studied. Solid endosperm and its carbonized version have a native grown distribution of light elements like K (from KCl), Na, Mg, etc. The hydrogen storage measurements reveal that the as synthesized materials have good hydrogen adsorption and desorption capacity with fast kinetics. The synthesized material adsorbs 2.30 wt. % at room temperature and 8.00 wt.% at liquid nitrogen temperature under 70 atm pressure. The present work is likely to initiate research on native light metal bearing carbon derived from natural precursors.

## 1 Introduction

It is generally agreed that the era of 'Hydrogen Economy' will solve the problem of depletion, pollution and global warming associated with fossil fuels. The missing link in harnessing hydrogen is 'storage'. The solid state storage in the form of hydride is the efficient and safe route of hydrogen storage. In order for hydrogen to become viable, certain specific limits on storage capacity, reversibility, cost effectiveness and abundant availability of the storage material are crucial aspects which must be fulfilled.

At present, built in hydride which we prefer to call as 'take out and fill in' variety of hydride typified by $MgH_2$ and complex hydride like $NaAlH_4$ are being extensively studied [1-6]. Some of these hydrides do meet the Department of Energy (DOE) 2015 criteria 5.5 wt. %, 40 gm/L [7]. However, attempts are being made to get rid of two main disadvantages of these hydrides namely high sorption temperature and slow kinetics. This is being done by catalyzing these hydrides through a variety of catalysts. However, one aspect seems to largely nullify the advantages of these hydrides. This is the lack of reversibility. Neither the elemental ($MgH_2$) nor the complex hydride ($NaAlH_4$ and other variants of these and similar hydrides) shows viable reversibility [7]. None of the studies seem to have shown the reversibility beyond few to about 100 cycles . These degrade to intolerable limits after about 100 cycles.

In view of the above, hydrogen storage in carbons, including porous carbon and nano structured carbon are being studied extensively [8-13].

One reason for this is that carbons show complete reversibility [7]. The other reasons for studies on carbons are that these are parts of nature's cycle, light weight, readily available through natural precursors, high surface area and tailorability of the texture (structure / microstructure).

The aim of the present study is to find out such type of carbon that shows good adsorption capacity with fast kinetics, particularly at ambient conditions, cost-effective and easily available. Carbon which can store hydrogen can be produced through a carbonization of a variety of materials. Some of these are polymers, coals, hydrocarbons biomass, fruits, etc. Out of variety of fruits, coconut is the one which has been widely investigated. These studies are mostly focused on the carbon produced from the coconut shell [14,15]. It may be mentioned that beside the coconut shell, there is a very useful biomass within the coconut. This is the solid endosperm which is also named as Kernel [16]. This biomass besides carbon also contains fatty acids, organic acids, more importantly it contains several light elements. Some of these are potassium, magnesium, sodium, calcium, etc [16,17] and some others are in lower concentration. Therefore, the carbon derived from this component of coconut is very different from other types of carbon obtained from other parts of coconut. Research is, therefore, required to investigate the hydrogen storage characteristics of solid endosperm carbon. The present work is motivated from these considerations. However, carbon which store hydrogen through adsorption of hydrogen molecules show viable storage capacity only at liquid nitrogen temperature (77K) but very low storage capacity $\leq 1$ wt. % at ambient conditions [7]. Thus it is important to carry out studies to improve the storage capacity of the carbon, particularly at room temperature. Most of the efforts on improving the storage capacity in carbon are focused on improving textural properties particularly surface area. However, it has now been realised that improvement in textural properties alone may not make carbon a viable hydrogen storage material [18]. It is also suggested that carbon derived from biomass may be an attractive form of carbon for hydrogen storage. The spillover effect, which may lead to higher hydrogen storage capacity, should also be investigated in addition to textural properties [7]. Keeping the above said aspects in view, we investigated carbons derived from a variety of biomass fruit precursors. It may be mentioned that, the storage capacity of carbons, including porous and activated carbons can be increased by doping/admixing with metallic element, including light elements [7]. We, therefore, selected solid endosperm part of biomass of the coconut as the precursor for making carbon. It is not only abundantly found but is also cost-effective and can be converted into carbon easily. Unlike graphene, Carbon nano tubes (CNTs) and other type of carbons, the production of carbon from solid endosperm is not time taking. Therefore, its cost effectiveness, good adsorption capacity and easy availability are its advantageous factors in regard to hydrogen storage.

## 2 Experimental details

**2.1 Sample Preparation:** As is known, that first step in the synthesis of carbon from biomass is carbonization. Samples were prepared by carbonization of dry solid endosperm part of coconut. The carbonization was done by heating the solid endosperm coconut at temperature 350 °C for 4 hours under nitrogen ($N_2$) atmosphere. Some samples have also been carbonized by heating at 600 °C. During this heating volatile material from the pores and moisture content got vaporized and removed.

It may be pointed out that for coconut derived carbon, comparatively lower carbonization temperature eg. 250 °C has been found to be best from the point of view of hydrogen storage characteristics [19]. Since the carbonized dry endosperm already contains pores, no special treatment was done to create pores.

**2.2 Characterizations and hydrogen adsorption, desorption studies**: Samples were characterized by using Scanning electron microscope (SEM) FEI Quanta 200, Transmission electron microscope (TEM) FEI: Technai $20G^2$ and X-ray diffraction (XRD) X' Pert PRO, PANalytical. For hydrogen sorption (adsorption/ desorption) studies, we utilized an automated-four channel Sievert's type apparatus; pressure composition isotherm (PCI unit, Advanced Materials Corporation, USA). In order to find out sorption characteristics samples were loaded into the sample chamber. Quartz wool is inserted in the sample chamber so that material could not damage the instrument during the evacuation. A vacuum level up to $10^{-6}$ torr was created. The pressure compression isotherms were then measured.

**2.3 Attempts of surface area measurement:** One rather anomalous feature of the porous carbon material regarding is the Brunauer, Emmett and Teller (BET) surface area. Typical BET surface area found by nitrogen adsorption at the surface is $\geq 10$ m$^2$/gm. This may be explained based on the fact that BET surface is revealed by adsorption of nitrogen molecules on the surface of carbon material. The BET surface area is derived on the basis of PCI isotherms measurement for the adsorbed $N_2$ molecules and their fitting with one of the three standard isotherms [20]. It may be pointed out that PCI isotherm may exhibit features not covered by the three standard isotherms. For such cases surface area cannot be evaluated based on BET technique [20]. In the present case, it may be noted that the surface contains multilayer macropores. Besides this whole material, including surface also embodies KCl tiny crystal and also several natural elements in the carbon derived from solid endosperm of coconut. These are KCl, Mg, etc. It may be pointed out that the finding of several elements is in keeping with the known result of the presence of these elements in dry coconut meat.

Because of the above said reason the surface area of the present carbon will not be of the type conducive for nearly uniform adsorption of nitrogen molecules. Thus the required condition for surface area through nitrogen BET technique is not satisfied here. In view of this, the low surface area observed in the present investigation can be understood. It may be pointed out that there have been cases where carbon with low surface area has also shown viable hydrogen storage capacity [21].

**3 Results and discussion**

**3.1 Hydrogen adsorption and desorption analysis of carbon derived from solid endosperm of coconut:** Fig 1 (a) shows the hydrogen storage capacity of 1.10 wt.% and 2.30 wt.% at room temperature and at 30 & 70 atm pressure respectively. Fig. 1 (b) shows representative hydrogen storage at liquid nitrogen temperature (77 K). The storage capacity is found to be 3.50 wt. % at 30 atm and 8.00 wt. % at 70 atm pressure. It may be pointed that the desorbed gas was checked through the residual gas analyser (SRS 200) attached to P-C-I apparatus. It was found that the desorbed gas is $H_2$. No trace of OH, $N_2$ and any other gas were found. The storage capacities are measured on several specimens and are found to be reproducible. It may be pointed out that our P-C-I system has been calibrated with standard sample $LaNi_5$ (capacity 1.50 wt. %). Also we are regularly measuring hydrogenation / dehydrogenation characteristics in catalysed $MgH_2$, $NaAlH_4$, $Mg(AlH_4)_2$ and several other complex hydrides. Thus the above result can be taken as reliable. It was confirmed by these results that material is adsorbing good amount of hydrogen. In carbon materials, a linear relationship exists between hydrogen storage capacity and pressure at room temperature [7]. This nature of the material is confirmed by PCI adsorption desorption curve. A typical PCI adsorption-desorption is shown in Fig. 1 (c). No hysteresis has been observed between adsorption and desorption curves. This confirms complete reversible nature of the present carbon derived from solid endosperm (CDSE) of coconut, like other carbon adsorbents [22].

**3.2 Structural/micro structural characterization of CDSE of coconut:** As it is given in the previous section (3.1) CDSE of coconut exhibit hydrogen storage capacity higher than conventional carbon, including porous carbon. These carbons show storage capacity <1 wt.% at room temperature and 4.0 to 5.0 wt.% at liquid nitrogen temperature [7]. It may, however, be pointed out that in some cases higher storage capacity at room temperature in carbon has also been reported [23]. These cases are for metal doped carbons where spillover effect is operative, and also in carbons prepared in a special way & hydrogen storage capacity measured at higher ($\geq$ 100) atmospheric pressure. Thus, the present results are curious. In order to find out the reason for it, we carried out detailed structural and microstructural characterization.

**3.3 X- ray diffraction study:** Fig.2 (a) and 2 (b) shows the XRD of CDSE of coconut before and after hydrogenation. It can be seen from these figures that broad XRD peak present at $2\theta = 19.9$ which can be indexed from carbon. However, in addition to the expected carbon XRD peak marked as '#' several other peaks labelled as '*' are also present. XRD pattern taken from 10 different samples prepared in different carbonization runs showed the ' #', '*' peaks. This implies that beside carbon, there is some other material embodied in it. Careful analysis showed that these peaks are explicable based on KCl cubic lattice structure with lattice parameter a = 6.293 A°. The presence of KCl together with carbon is quite curious. We shall discuss this in the forth coming section.

**3.4 TEM structure of the sample:** Fig.3 shows a typical TEM micrographs obtained from the CDSE of coconut. It can be seen that the TEM micrograph exhibited a secondary phase. Selected area electron diffraction pattern (Fig.3(a)$_{D.P}$)from the main matrix showed two halos like diffraction rings. The most intense ring corresponds to d= 3.43 A°. This shows the presence of disordered carbon matrix. It is in keeping with the XRD results. The selected area diffraction patterns taken from the precipitate like regions Fig.3(b),Fig.3(c) exhibited spot diffraction pattern. Representative spot diffraction patterns are shown in (Fig. 3(b)$_{D.P}$ & 3(c)$_{D.P.}$) Indexing of these diffraction patterns reveled that it could be indexed successfully based on Mg and KCl lattice. It may be pointed out that the coconut solid endosperm is known to have native as grown elements, compounds like K, Na, Mg, Cl, fatty acids, etc.

In order to identify the elements present in the sample EDX spectra was taken. Fig.4 shows the respective EDX spectra of the sample. It can be seen from the table.1,that elements present are dominantly carbon, followed by potassium, magnesium, molybdenum, chlorine, phosphorus, etc. The elements, Mg, Mo, P, etc. could not be detected by XRD. This may be due to their small concentration.

**3.5 SEM exploration by the sample of CDSE of coconut:** In view of the curious XRD and TEM results of CDSE, we carried out further exploration of the sample by SEM with particular attention of the other phases / elements and the texture of pores in the carbon matrix. Fig 5 (a), 5(b) shows the layered and porous structure of the CDSE of coconut and Fig. 5 (c) which is a high magnified SEM microstructure, shows that there is a nearly uniform distribution of the secondary phases embodied in the main carbon matrix. The presence of tiny regions can be clearly seen in this figure. These secondary phases which are native built in regions are, in view of XRD and EDX results, can be taken to be KCl.

**.4 Mechanism for enhanced hydrogen storage capacity of CDSE:** As given in the section 3.3 & 3.4, the XRD and TEM results clearly show that CDSE of coconut invariably coexists with KCl. It should be pointed that coconut composition, including solid endosperm depends on the place where it grows. Different concentration of elements and compounds has been reported. Thus in one report besides fatty acids, vitamins, the elemental concentration given is K~ 600 mg, Mg~ 46 mg, Na~ 6 mg, Ca~2 mg, in another report, the relative abundance of the element given is K~ 370 mg, Na~ 260 mg. In yet another place there are K~ 312 mg, Mg~ 30 mg, Na~ 105 mg, Cl~ 183 mg are reported (100 gm is reference level for all) [24]. It may be pointed out that the above said concentration is for as grown solid endosperm of coconut. There is no report for carbonized version of CDSE except the one given in the present work.

It may be pointed out that even though the melting /evaporation temperature of KCl in carbon matrix is not known, this temperature for the bare compound is quite high (770 °C) for KCl.

From the above it can be taken that the KCl phase is embodied and is invariably present in the CDSE of coconut. The difference in electro negativity of K and Cl is 3.0- 0.8 = 2.2 and hence KCl is a polar molecule. Since KCl is embodied in carbon matrix, the presence of KCl will polarize the carbon matrix surrounding it. When a carbon molecule is adsorbed on polarised carbon matrix, it will be adsorbed with higher bond strength than in a non polarized neutral matrix. It is known that the binding energy of adsorption on carbon is 1–8 KJ mol$^{-1}$ [7]. This is too less for reasonable storage capacity at room temperature. However, it has been known although not experimentally realised that in a polarised carbon matrix, the physisorption will have higher binding energy of adsorbed hydrogen molecules [23]. This will enhance the hydrogen adsorption capacity. Although it is not possible to evaluate the binding energy of adsorbed hydrogen for the present polarized carbon matrix, it is clear that hydrogen adsorbed in polarized matrix will have higher bond strength. Therefore, the corresponding normal carbon configuration at room temperature, as observed in the present study, 1.10 wt% at 30 atm and 2.30 wt.% at 70 atm respectively. Because of the polarization effect, the enhanced binding energy of adsorbed $H_2$ on carbon will get added to the enhancement of binding energy resulting from lower temperature ( 77 K). In view of this higher hydrogen storage capacity of CDSE viz 3.00 wt.% at 30 atm and 8.00 wt.% at 70 atm at liquid nitrogen temperature can be understood.

As regards other elements present in CDSE, Mg is of interest since it is capable of working as a dissociation source for hydrogen molecule. Since the elements present in CDSE come from natural / native dispersion effect, they are expected to be more uniformly distributed. Under these conditions the hydrogen atoms coming from dissociation of $H_2$ molecules, will get spilled over to the receptor in the carbon layers of CDSE. Although the spillover effect is not well understood, the effect is well recognised as one potential sources of enhancement of hydrogen storage capacity [23]. The presence of Mg suggests that the enhancement in hydrogen storage capacity in the present CDSE may be at least partly due to spill over effect.

**Conclusions:** The present work elucidates that carbon derived from solid endosperm of coconut exhibits considerable hydrogen storage capacity of 2.30 wt. % at room temperature and 8.00 wt.% at liquid nitrogen temperature, at 70 atm pressure. It has been suggested that this comparatively high storage capacity arises due to the presence of native built in KCl phase embodied in the carbon matrix. Another reason for the high storage capacity may be due to the spillover effect produced by built in elements such as Mg, capable of dissociating the hydrogen molecule and transfer of the hydrogen atoms in the carbon layers derived from solid endosperm. The present study indicates that hydrogen storage in such carbons, which have native as grown compounds and elements need to be further explored.

**Acknowledgement**
The authors would like to thank Dr. V.Sekkar, Prof. R.S. Tiwari, Dr. M.A. Shaz, Dr. Pragya jain, Miss. Sunita K Pandey for their keen interest and useful discussions. The authors would also like to thank University Grant Commission (UGC) (India) and Department of Science and Technology (DST) (India) for their support.

**Figure caption:**
**Fig 1 :**(a) Hydrogen adsorption kinetics at room temperature and (b) at Liquid nitrogen temperature. (c) Pressure Composition Isotherm (PCI) of CDSE of coconut at room temperature.
**Fig 2:** XRD pattern of CDSE of coconut: (a) before hydrogenation and (b) after hydrogenation. The carbon peaks are marked as '#' and that from KCl as '*'. It may be notice that KCl remains present even after hydrogenation.
**Fig 3:** TEM micrograph of CDSE of coconut: (a) diffraction rings in the form of halos show the presence of disordered carbon matrix (b) hexagonal diffraction pattern shows the presence of Mg (c) diffraction pattern shows the presence of KCl.
**Fig 4:** EDX of CDSE of coconut shows the presence of Mg, P, Mo, Cl, K.
**Fig 5:** SEM images (a) & (b) show the layered and porous nature of CDSE of coconut. Besides these as shown in fig. (c) built in native region of KCl in the form of tiny particles are also present (see text for details).

**Fig.1**

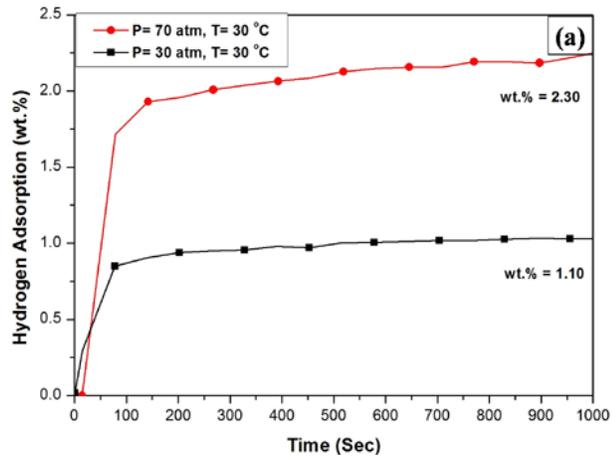

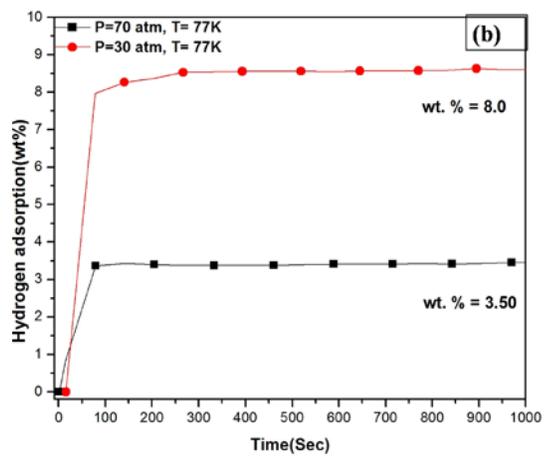

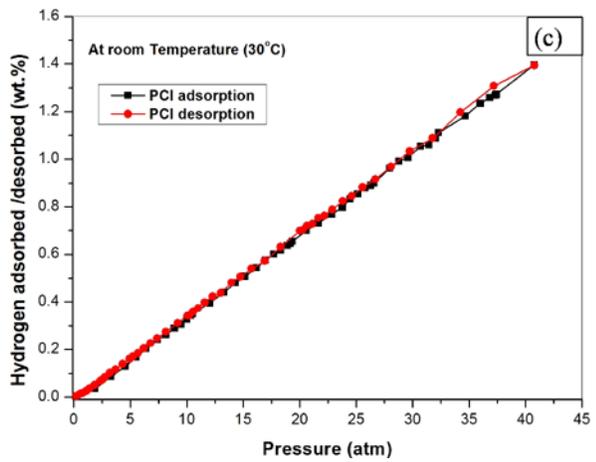

**Fig.2**

**Fig.3**

**Fig.4**

| Element | Weight % | Atomic % |
|---|---|---|
| C (K) | 70.60 | 82.60 |
| Mg (K) | 01.20 | 00.70 |
| P (K) | 02.90 | 01.30 |
| Mo (L) | 03.50 | 00.50 |
| Cl (K) | 01.00 | 00.40 |
| K (K) | 07.20 | 02.60 |

**Fig.5**

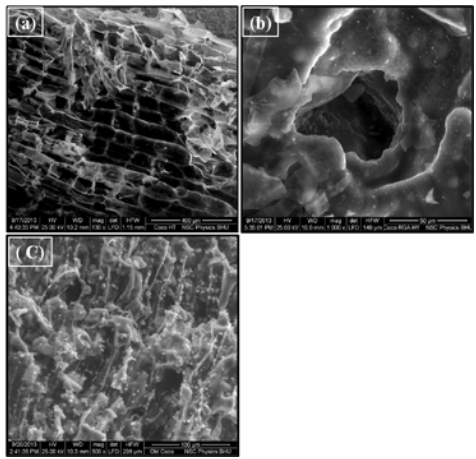